# Saturated Ferromagnetism and Magnetization Deficit in Optimally Annealed Ga$_{1-x}$Mn$_x$As Epilayers

S. J. Potashnik, K. C. Ku, R. Mahendiran, S. H. Chun, R. F. Wang, N. Samarth, and P. Schiffer[*]

*Department of Physics and Materials Research Institute, Pennsylvania State University, University Park, PA 16802*

**Abstract**

We examine the Mn concentration dependence of the electronic and magnetic properties of optimally annealed Ga$_{1-x}$Mn$_x$As epilayers for $1.35\% \leq x \leq 8.3\%$. The Curie temperature ($T_c$), conductivity, and exchange energy increase with Mn concentration up to $x \sim 0.05$, but are almost constant for larger $x$, with $T_c \sim 110$ K. The ferromagnetic moment per Mn ion decreases monotonically with increasing $x$, implying that an increasing fraction of the Mn spins do not participate in the ferromagnetism. By contrast, the derived domain wall thickness, an important parameter for device design, remains surprisingly constant.

[*] e-mail: schiffer@phys.psu.edu

Ferromagnetic semiconductors are of great current interest [1,2] since they enable the incorporation of ferromagnetic elements into important electronic device configurations, and they are also expected to be essential for efficient spin injection into non-magnetic semiconductors [3,4,5,6]. One of the most promising model systems in which to study ferromagnetism in a semiconductor is $Ga_{1-x}Mn_xAs$, which can be grown epitaxially within III-V heterostructures. $Ga_{1-x}Mn_xAs$ has a ferromagnetic ground state for $x \gtrsim 0.015$, and the maximum reported value of the ferromagnetic transition temperature is $T_c \sim 110$ K [7]. Despite the importance of this material as a model system, the fundamental physics underlying its physical properties remains the subject of much theoretical speculation [8,9,10,11,12,13,14,15].

The complexity of the ferromagnetism in $Ga_{1-x}Mn_xAs$ originates in the interplay between the magnetic and electronic properties associated with the Mn dopants. The spin 5/2 Mn ions act as acceptors, providing holes that mediate a ferromagnetic Mn-Mn coupling [16], but these holes are heavily compensated by defects in the material (which is grown by low-temperature molecular beam epitaxy (MBE)) [17]. The importance of these defects has been demonstrated by experimental studies which show that -- even for a fixed Mn composition -- the physical properties are highly sensitive to the detailed MBE growth conditions [18,19]. Consistent with this finding, post-growth annealing alters the defects and can either enhance or degrade the ferromagnetic properties (e.g. the total moment and $T_c$) [20,21,22]. An optimal anneal (1-2 hours at $T \sim 250$ C) can maximize both the conductivity and $T_c$ for a given sample up to the highest value reported in the literature, $T_c \sim 110$ K [22].

Due to the sensitivity of $Ga_{1-x}Mn_xAs$ to growth conditions, it is quite difficult to compare different samples and obtain a clear picture as to how the ferromagnetism evolves as a function of Mn concentration ($x$). The generally accepted compositional phase diagram [7] shows $T_c$ increasing monotonically to $x \sim 0.05$, and then decreasing for larger $x$. We have examined a series of optimally annealed samples of $Ga_{1-x}Mn_xAs$ which were grown in a continuous sequence with increasing Mn concentration. We find that $T_c$, the conductivity, and the ferromagnetic exchange energy increase with Mn concentration for low $x$, but they all become nearly independent of $x$ for $x > 0.05$, where $T_c$ reaches the established maximum value of 110 K. In this range of large $x$, the temperature dependence of the magnetization and $T_c$ can be described within a conventional 3D Heisenberg model. By contrast, the ferromagnetic moment per Mn ion decreases monotonically with $x$ throughout the entire range of doping, reflecting that a large fraction of Mn ions do not participate in this conventional ferromagnetic state.

We studied a series of ferromagnetic $Ga_{1-x}Mn_xAs$ samples with $0.0135 \leq x \leq 0.083$, all of which were grown in a continuous series of increasing Mn content (except for $x = 0.083$, grown under similar conditions in a previous run [22]). The samples were grown on (100) semi-insulating, epiready GaAs substrates under conditions which have been described previously [22]. The continuous series of $Ga_{1-x}Mn_xAs$ epilayers were $123 \pm 2$ nm thick and grown on a buffer structure consisting of a standard (high temperature grown) 100 nm GaAs epilayer followed by a 25 nm low temperature grown GaAs epilayer which leads to compressive strain and hence the easy orientation of the magnetization being in-plane. Fringes seen in high resolution x-ray diffraction show the samples to be high quality. Samples were annealed for 1.5 hours at 250°C in a 99.999% purity flowing nitrogen atmosphere. Magnetization was measured in-plane with a commercial superconducting quantum interference device magnetometer in a field of 0.005 T after cooling in a 1 T field. High resolution x-ray diffraction measurements were performed with a Philips four circle diffractometer with λ =1.54 Å Cu Kα x-rays provided by a fixed source tube fitted with a double bounce monochromator. X-ray and magnetization data taken to $T > 320$ K show no evidence of MnAs precipitates, although we cannot exclude the possibility of nanoscale MnAs clusters which would be superparamagnetic near the bulk MnAs $T_c$. Samples in the continuous series were grown in the order of increasing Mn source temperature which yielded increasing Mn concentrations. The exact Mn concentrations of the annealed samples were determined by electron microprobe analysis (EMPA) that shows a monotonically increasing $x$ with the Mn source temperature as expected (figure 2 inset); details of the EMPA analysis are described elsewhere [22].

In figure 1 we show the typical temperature dependence of the resistivity and magnetization which is qualitatively similar to that in the "colossal magnetoresistance" manganites [23], although the mixed Mn valence associated with double exchange in those systems is not present in $Ga_{1-x}Mn_xAs$. The resistivity follows a variable range hopping form above $T_c$ [24], but it reaches a maximum at $T_c$ and then decreases as the temperature is lowered further. The magnetization, $M(T)$, displays a sharp rise at $T_c$, as expected for long range ferromagnetic order. The linear increase and kink in $M(T)$ below $T_c$ is generally observed in as-grown samples, and has been attributed to multiple exchange interactions [21] or non-collinear ferromagnetism [25]. By contrast, the optimally annealed samples display a much more conventional ferromagnetic $M(T)$ at low temperatures [22] as discussed below.

We now discuss how the physical properties of $Ga_{1-x}Mn_xAs$ evolve with Mn content. As shown in figure 2, the conductivity at $T = 300$ K increases with increasing Mn concentration, consistent with the increased doping represented by the Mn ions, although there is not a significant increase in the conductivity above $x \sim 0.05$. In figure 3a, we plot the concentration dependence of $T_c$ (as determined by the onset of a ferromagnetic moment in M(T)) for the annealed samples. As reported in previous studies (on as-grown samples), $T_c$ increases with $x$ for $x < 0.05$. For $x > 0.05$, however, $T_c$ in the annealed samples reaches 110 K and becomes independent of Mn concentration up to our highest value of $x$. We note that this concentration dependence of $T_c$ in the annealed samples is paralleled by that of the lattice spacing which is consistently smaller than that of the unannealed samples (Fig. 3(b)) [26]. The $T_c$ and conductivity data define a compositional phase diagram of $Ga_{1-x}Mn_xAs$ which is quite different from that reported in reference [7], where a single sample at $x = 0.071$ suggested that $T_c(x)$ decreased dramatically at large $x$ with a corresponding increase in the resistivity. By contrast, we find that $T_c$ does not decrease and the samples do not become less metallic as $x$ exceeds 0.05, but rather that both of these properties saturate at large $x$. This saturation could possibly be explained by an increasing fraction of Mn ions not participating in the ferromagnetism, as described below.

Our measurements of $M(T)$ also allow us to obtain the exchange energy $J$ associated with the ferromagnetic state. As shown by the solid line fit in figure 1, we can understand $M(T)$ in the low temperature limit within a standard three-dimensional spin wave model which predicts $M(T) = M_0 - 0.117\mu_B(k_BT/2SJd^2)^{3/2}$ where $M_0$ is the zero temperature magnetization, $d$ is the spacing between Mn ions, and $J$ is the exchange interaction [27]. Within a three-dimensional Heisenberg model with nearest neighbor interactions $J$ is related to $T_c$ through the relation $J = 3k_BT_c/(2zS(S+1))$ where z is the number of nearest neighbor spins. Since $z$ is not well defined in this random system and the interactions are long-range, we somewhat arbitrarily treat the Mn lattice as cubic, i.e. $z = 6$ (thus introducing an uncertainty of perhaps a factor of 2). We then compare values of $J$ obtained through the two different methods in figure 3c. As would be expected from our values of $T_c$, $J$ increases with $x$ up to $x \sim 0.05$, but remains approximately constant for larger $x$. The values of $J$ are somewhat smaller than the theoretical prediction of König *et al.* [10], but the agreement within a factor of two is reasonable given the approximate nature of both calculations [28]. Importantly, we also find good agreement between the two methods of estimating the exchange energy, especially at larger values of $x$. This agreement strongly indicates that ferromagnetism in $Ga_{1-x}Mn_xAs$ can be understood within a conventional model, and also substantiates the Mn-concentration independence of the ferromagnetism in the high $x$ regime since it is reflected both in $T_c$ and in the low temperature limiting magnetization.

From the fits to $M(T)$ we also extract the zero temperature magnetization ($M_0$) which corresponds to the number of spins in the ferromagnetic state. As shown in figure 4a, $M_0$ increases with Mn concentration as might be expected. As has been noted by other authors [1,21,29], the moment per Mn ion (figure 4b) is well below the full saturation value of $5\mu_B$ for spin 5/2 moments. We find that this normalized moment decreases with increasing $x$ throughout the entire range of doping, falling to about half of the expected moment for all the Mn spins at large x (although the observed moment is always much larger than that inferred from circular dichroism measurements at $x = 0.02$ [29]). This magnetization deficit implies that, as the concentration of spins in the material increases, a decreasing fraction are actually participating in the ferromagnetism.

It is important to understand the magnetization deficit, since if one could induce more spins to participate in the ferromagnetism, there might be a continuing increase in $T_c$ beyond 110 K. One possible explanation for the deficit is that some fraction of the spins form a spin-glass-like state at low temperatures in parallel with the ferromagnetism. We find, however, no irreversibility between field-cooled and zero-field cooled data at 0.01 T. Another possible explanation is that the ferromagnetic state is homogeneous but non-collinear [25], although the conventional behavior of $M(T)$ in the annealed samples would argue against this explanation. Additionally, applied fields up to 7T (which should quench a spin glass state and align a non-collinear state) do not reveal additional moment approaching the magnitude of the deficit. A more likely explanation is that the local electronic structure associated with certain defects (such as nanoscale MnAs clusters or Mn ions forming six-fold coordinated centers with As [21] as has been explicitly observed in $In_{1-x}Mn_xAs$ [30]) precludes individual Mn moments from participating in the ferromagnetism \* MERGEFORMAT . Our data would be consistent with an increase in the density of such defects with increasing Mn content and also the observed strong dependence of $M_0$ on annealing time [22].

The domain structure of $Ga_{1-x}Mn_xAs$ is also of considerable interest [31, 32], and our fits to the magnetization data also allow us to calculate the domain wall thickness ($t$). We model the domain wall as a two dimensional Néel wall with no in-plane anisotropy [29] (verified by coercive field measurements on our samples) which should have a thickness of the form $t = \{\pi JS^2/[2(M_0)^2 d]\}^{1/2}$ [33]. From our above determination of $d$, $J$, $M_0$, we get an average domain wall thickness of about 17 nm which is remarkably constant over the entire range of Mn concentrations studied, as shown in figure 4c. This thickness, which has not been measured experimentally, would not be inconsistent with the measured domain size of 1.5 μm for $Ga_{0.957}Mn_{0.043}As$ [31]. A thin domain wall is attractive for applications such as the spin diode and spin transistors [6] as well as possible spin filters [35] because the spin of the carriers traversing the wall will be relatively unaffected, and its insensitivity to Mn concentration will facilitate device fabrication.

In summary, we have studied the compositional dependence of the magnetic properties of $Ga_{1-x}Mn_xAs$ in the ferromagnetic regime. We find that $T_c$, the exchange energy, and the conductivity all saturate for large values of $x$ and that the ferromagnetism at large $x$ is quite conventional, even though the saturation moment is well below theoretical expectations and decreases with increasing Mn content. These findings place new constraints on theoretical explanations for ferromagnetism in this important model system.

The authors would like to thank A. Millis, M. Flatté, and A. MacDonald for valuable discussions. N.S., S.H.C., K.C.K. were supported by ONR N00014-99-1-0071 and –0716, and DARPA/ONR N00014-99-1-1093. S.J.P., R.M., R.F.W. and P.S. were supported by DARPA N00014-00-1-0951 and NSF DMR-0101318.

# Figure Captions

**Figure 1.** The magnetization and resistivity as a function of temperature for $Ga_{1-x}Mn_xAs$ with $x = 0.0597$ (circles, as-grown; triangles, annealed). The thick solid line is a fit to the magnetization data for $T \leq 40$ K as described in the text.

**Figure 2.** The conductivity at 300 K as a function of Mn concentration. The conductivity is seen to increase with increasing Mn concentration up to $x \sim 0.05$ and remains approximately constant for larger $x$. The inset shows the Mn content as determined by electron microprobe analysis as a function of Mn source temperature.

**Figure 3. a)** and **b)** The transition temperature ($T_c$) and the relaxed lattice constant as a function of Mn concentration for the optimally annealed samples (solid circles). Data from unannealed samples (open circles) are shown for comparison. Previous measurements of the lattice constant of unannealed samples are shown by the solid [19] and dashed lines [1]. **c)** The exchange energy ($J$) as a function of Mn concentration. The solid circles represent values obtained from fits to $M(T)$ and the open triangles represent values determined from $T_c$ as described in the text.

**Figure 4. a)** and **b)** The zero-temperature magnetization ($M_0$) and the magnetization per Mn atom as a function of Mn concentration ($x$) for the optimally annealed samples. Note that the latter decreases monotonically with $x$ for the entire range of samples studied. **c)** The estimated domain wall thickness for the range of samples studied.

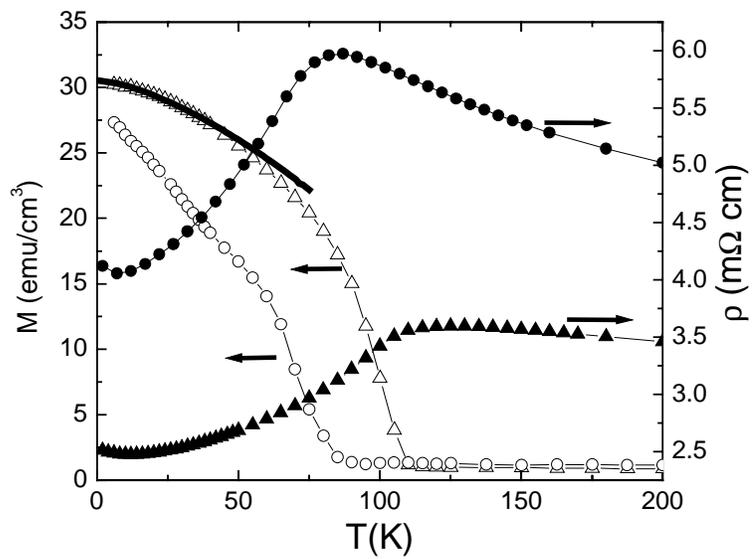

Potashnik et al. -- Figure 1

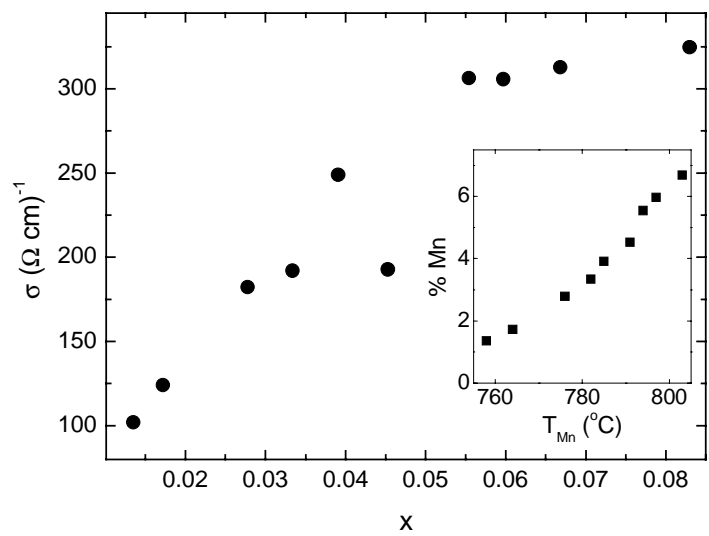

Potashnik et al. -- Figure 2

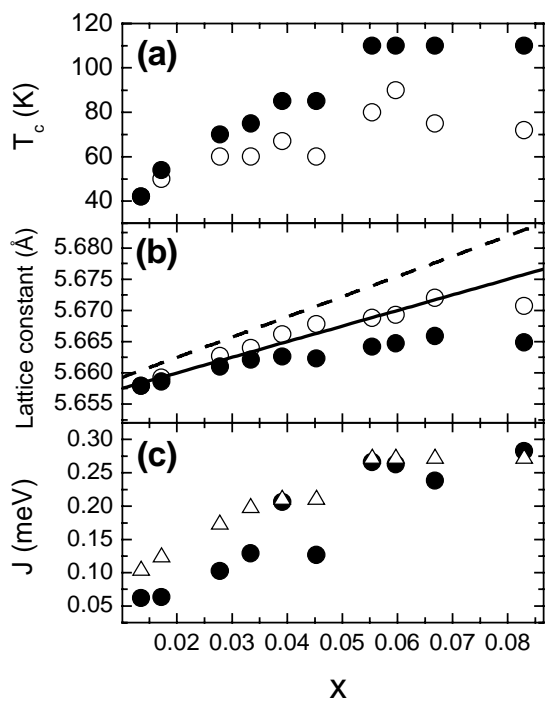

Potashnik et al. – Figure 3

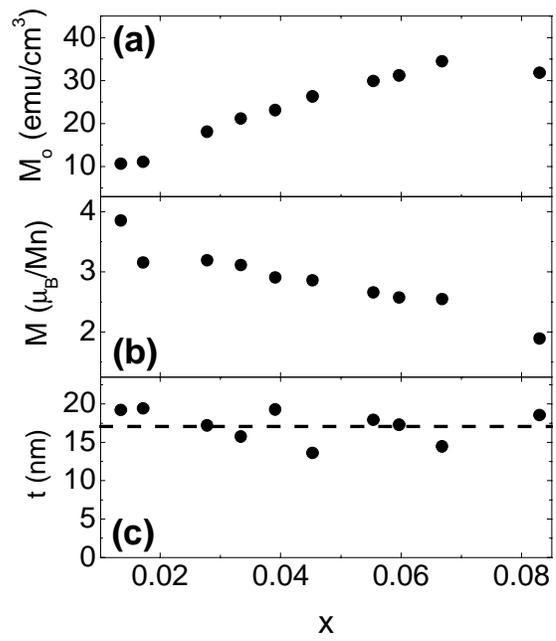

Potashnik et al. -- Figure 4